\documentclass[12pt,fleqn,leqno]{article}
\usepackage{amsmath,amssymb}
\usepackage{multicol}
\usepackage{amsthm}
\usepackage[dvips]{graphicx}

\allowdisplaybreaks

\numberwithin{equation}{section}

\makeatletter
\renewcommand{\section}{\@startsection{section}{1}{0pt}{20pt}{6pt}{\large\bf}}
\renewcommand{\@seccntformat}[1]{\csname the#1\endcsname.\ }

\def\footnoterule{\kern -3pt \hrule width 2.7 true cm \kern 2.6pt}

\def\ni{\noindent}
\def\vs{\vspace}
\def\hs{\hspace}

\def\EE{\mathsf E}
\def\PP{\mathsf P}

\def\R{I\!\!R}
\def\L{I\!\!L}
\def\wt{\widetilde}
\def\wh{\widehat}

\def\e{\text{e}}

\newcommand{\p}{\! +\! }
\newcommand{\m}{\! -\! }

\newtheorem{theorem}{Theorem}[section]

\newtheorem{remark}[theorem]{Remark}

\begin{document}

\title{\textbf{On the American swaption in the linear-rational framework}\footnote{The authors wish to thank Anders Trolle for his comments and for providing the parameters for the numerical example. We also thank the referee for useful comments that helped to improve the paper. The research leading to
these results has received funding from the European Research Council under the European UnionÕs Seventh
Framework Programme (FP/2007-2013) / ERC Grant Agreement n.\ 307465-POLYTE.}}
\author{Damir Filipovi\'c\thanks{EPFL and Swiss Finance Institute, Lausanne, Switzerland. \texttt{damir.filipovic@epfl.ch}}\:\:\:\: \emph{and} \:\:\:\:Yerkin Kitapbayev\thanks{Questrom School of Business, Boston University, Boston, MA, USA; \texttt{yerkin@bu.edu}}
}
\maketitle


{\par \leftskip=2.6cm \rightskip=2.6cm \footnotesize

We study American swaptions in the linear-rational (LR) term structure model introduced in \cite{FLT}. The American swaption pricing problem boils down to an optimal stopping problem that is analytically tractable. It reduces to a free-boundary problem that we tackle by the local time-space calculus of \cite{Pe-1}. We characterize the optimal stopping boundary as the unique solution to a nonlinear integral equation that can be readily solved numerically. We obtain the arbitrage-free price of the American swaption and the optimal exercise strategies in terms of swap rates for both fixed-rate payer and receiver swaps. Finally, we show that Bermudan swaptions can be efficiently priced as well.

\par}




\vs{-18pt}

\vs{-18pt}

\section{Introduction}

An interest rate swap is a contract between two parties who agree to exchange cash flows over a pre-specified time grid. The holder of a payer (receiver) swap contract pays a fixed (floating) rate and receives a floating (fixed) rate on a notional amount. The floating rate is usually tied to the London Interbank Offered Rate (LIBOR), a daily fixed standard benchmark for interest rates in various currencies. A payer (receiver) swaption gives the holder the right but not the obligation to enter a payer (receiver) swap at a pre-specified fixed strike rate. Swaptions form an important class of derivatives that allow to price and hedge interest rate risk. They underlie callable mortgage-backed securities, life insurance products, and a wide variety of structured products. The outstanding notional amount in the swap market is in the order of hundreds of trillions US dollars.

Swaptions can be divided into three classes according to their exercise timing rights. European swaptions can be exercised only at the expiration date. American swaptions allow the holder to enter the swap on any date that falls within a range of two dates. Bermudan swaptions constitute a simplified variant of American swaptions where exercise is only possible on a finite time grid.
The pricing of American swaptions is arguably a difficult task. The standard approach is to approximate American by Bermudan swaptions and price the latter using simple tree models or Monte-Carlo simulation based methods (see e.g. \cite{LSS}). To our knowledge there has not been a model in the literature for which American swaptions are priced analytically in continuous time. A reason why American swaption pricing in continuous time was not feasible so far is that in most interest rate models used in the literature before \cite{FLT}, the payoff of a swaption is the positive part of a sum of exponential-affine functions in the factor. This cannot simply be reduced to an optimal stopping problem for the factor process.

In this paper we analytically price American swaptions in the one-factor linear-rational term structure model introduced in \cite{FLT}. The idea of the model and important property that we exploit is that discounted bond prices are linear in the factor. In order to fit the term structure of European swaptions we allow for time-varying diffusion coefficient of the factor process.
Using this we obtain that the American swaption pricing problem boils down to an undiscounted optimal stopping problem for a scalar diffusion process. The latter problem is reduced to a free-boundary problem that we tackle by the local time-space calculus of \cite{Pe-1}. We characterize the optimal stopping boundary as the unique solution to a nonlinear integral equation that can readily be solved numerically. Using these boundaries we obtain the American swaption price and the optimal exercise strategies in terms of the swap rate for both payer and receiver swaptions.
We provide the numerical algorithm for solving integral equations for the boundaries and computing the American swaption prices.
The numerical example in Section~\ref{secnumer} illustrates this algorithm and we solve the American swaption pricing problem for the calibrated set of parameters. We also
show that the Bermudan swaptions can be efficiently priced under the linear-rational term structure model (Section~\ref{Bermudan}). We then compare the prices of
European, American and Bermudan swaptions. The discrepancy between American and Bermudan prices is quite small. 


Another important application of this paper is based on the equivalence of the payoffs of a receiver (payer) swaption and a call (put) option on a coupon bearing bond. Hence our results cover the pricing of American options on coupon bearing bonds which is itself 
interesting practical problem. Also it is well known that American swaption can be used to convert non-callable bond to callable bond. Here is an example for illustration.
 Suppose a company has issued a bond maturing in 10 years with annual coupons of 4\% on the principal amount $N$ and wants to add the option to call (prepay) the bond at par (for $N$) at any time $\tau$ before maturity date. This option means that the company has the right to prepay the principal $N$ of the bond at time $\tau$ and stop paying coupons afterwards. If the company cannot change the original
bond, it could buy an American receiver swaption with strike rate 4\% on the swap with period $[0,10]$. If the company exercises the swaption at time $\tau\in[0,10]$, the fixed coupon leg of the swap will then cancel the fixed coupon payments of the bond. On the other hand, paying the floating rate leg of the swap and the principal $N$ of the bond at maturity $T=10$ is equivalent to paying $N$ at $\tau$, as desired. This is a modified version of Example 2.1 in \cite{F-1}.

The structure of the paper is as follows. Section~\ref{secmod} introduces the linear-rational term structure model and  translates the American swaption pricing problem to an optimal stopping problem. Section~\ref{secpay} reduces the optimal stopping problem to a free-boundary problem for American payer swaptions. Section~\ref{secrec} does the same for American receiver swaptions. Section~\ref{secnum} provides an algorithm for numerically solving integral equations arising in the solution of the free-boundary problem. Section~\ref{secoeb} expresses the optimal exercise strategies for American swaptions in terms of the underlying swap rate. Section~\ref{Bermudan} discusses the pricing of Bermudan options under the linear-rational framework.
Section~\ref{secnumer} presents the numerical results.  Finally, Section~\ref{secconc} concludes and provide the agenda for future research.

\vs{6pt}

\section{Model setup and formulation of the problem}\label{secmod}

1. We consider the one-factor linear-rational square-root diffusion model introduced in \cite{FLT}. The factor process is a square-root diffusion $X$ given by
\begin{equation} \label{SDE} \hs{6pc}
 dX_{t}=\kappa (\theta \m X_{t})\,dt+\sigma(t) \sqrt{X_{t}}\,dB_{t}\;\;\; (X_0> 0)
 \end{equation}
where $B$ is a standard Brownian motion started at zero and $\kappa$ and $\theta$ are positive parameters, and $\sigma(t)$ is time-varying deterministic continuous function for $t>0$. A sufficient condition for the absence of arbitrage opportunities in a financial market model is the existence of a state price density: a positive adapted process $\zeta_t$ such that the price $\Pi(t,T)$ at time $t$ of any cash flow $C_T$ at time $T$ is given by
  \begin{align} \label{price-1} \hs{6pc}
 \Pi(t,T)=\frac{1}{\zeta_t}\EE_t [\zeta_T C_T].
 \end{align}
We specify the state price density as
\begin{equation} \label{state price} \hs{6pc}
\zeta_t=e^{-\int_0^t \alpha(s)ds}(1+ X_t)
 \end{equation}
 where the function $\alpha:[0,\infty)\mapsto \R$ is
a deterministic continuous function.

The main feature of the model \eqref{SDE}-\eqref{state price} is that it provides tractable expressions for zero-coupon bond prices $P(t,T)$
 with $C_T=1$
  \begin{align} \label{price-2} \hs{4pc}
 P(t,T)=\frac{\EE_t [\zeta_T]}{\zeta_t}=e^{-\int_t^T \alpha(s)ds}\frac{1+\theta+e^{-\kappa(T-t)}(X_t \m\theta)}{1+ X_t}
  \end{align}
 where we used the following conditional expectations
\begin{align} \label{cond-x} \hs{6pc}
 &\EE_t [X_T]=\theta+e^{-\kappa(T-t)}(X_t \m\theta)\\
 &\label{cond-zeta}
 \EE_t [\zeta_T]=e^{-\int_0^T \alpha(s)ds}(1+\theta+e^{-\kappa(T-t)}(X_t \m\theta))
 \end{align}
 for $0\le t\le T$. The function $\alpha$ can then be chosen such that the model-implied zero-coupon bond
prices exactly match the observed term structure at time $t=0$ (see Appendix G in \cite{FLT}). Formula \eqref{price-2} explains why this model is called linear-rational in \cite{FLT}.
 The short rate is obtained via the relation $r_t =- \partial_{T} \log P(t,T)_{|T=t}$ and is given by
\begin{align} \label{price-3} \hs{6pc}
r_t = \alpha(t) -\frac{\kappa (\theta \m X_{t})}{1+X_t}.
\end{align}
Consequently, it is bounded from below and above by
\begin{align} \label{srbounds} \hs{6pc}
\alpha(t)-\kappa\theta\le r_t   \le \alpha(t)+\kappa.
\end{align}

In \cite{FLT} this model has been studied thoroughly both analytically and numerically, especially from a swap and European swaption pricing point of view. In \cite{FLT} it is also shown that the bounds in \eqref{srbounds} are not economically binding. In this paper we consider the application of the linear-rational framework to the pricing of the American-style swaptions.

\vs{6pt}

2. Let us now introduce a fixed versus floating interest rate swap which is specified by a tenor structure of reset and payment dates
$0<T_0<T_1<...<T_n$, where we assume that $T_i -T_{i-1}=\Delta$ for $i=1,..,n$ to be constant, and a pre-determined fixed rate $K$. At each date $T_i$,
 $i=1,..,n$, the fixed leg pays
$\Delta\times K$ and the floating leg pays $\Delta\times$LIBOR accrued over the preceding time period. From the
perspective of the fixed-rate payer, the value of the swap at time $t\le T_0$ is then given by
 \begin{align} \label{payoff-1} \hs{6pc}
 \Pi^{swap}_t=P(t,T_0)-P(t,T_n)-\Delta K\sum_{j=1}^{n}P(t,T_j).
  \end{align}
  A payer swaption is an option to enter into an interest rate swap, paying the fixed leg at
a pre-determined rate and receiving the floating leg. A European payer swaption expiring
at $T_0$ on a swap with the characteristics described above has a value (payoff) at expiry $T_0$
\begin{align} \label{payoff-2} \hs{6pc}
 C_{T_0}=(\Pi^{swap}_{T_0})^+=\Big(1-P(T_0,T_n)-\Delta K\sum_{j=1}^{n}P(T_0,T_j)\Big)^+.
  \end{align}
  In the linear-rational framework the price of European payer swaption at time $t\le T_0$ with $X_t=x$ equals
\begin{align} \label{payoff-3} \hs{6pc}
 V^{E}(t,x)=\frac{1}{\zeta_t}\EE_{t,x} [\zeta_{T_0} C_{T_0}]=\frac{1}{\zeta_t}\EE_{t,x} [p (X_{T_0})^+]
 \end{align}
where the expectation $\EE_{t,x}$ is taken under condition that $X_t=x$ and $p(x)$ is an explicit linear function of $x$.
Throughout the paper, we will also use the notation $X^{t,x}_u$ for $u\ge t$ when  $X_t=x$.

 \begin{remark}
 The reason we allow for time-varying $\sigma$ in \eqref{SDE} is that we would like to fit the data of European swaption prices at current date $t$ for different maturities $T_0>t$ and swaption lenghts
 $T_n-T_0$.
It can be seen from \eqref{price-2} that $\sigma(t)$ does not effect ZCB prices. Therefore the calibration can be done as follows: we estimate $(\alpha,\kappa,\theta,X_0)$ to fit the set of  spot and forward swap rates, and then calibrate $\sigma(t)$ to match the European swaption prices.
 Unfortunately, time-varying $\sigma$ complicates the numerical analysis in this paper as the process $X$ becomes time-inhomogeneous and also the probability density function is not available explicitly anymore unlike in the case of constant $\sigma$ (CIR process). However, we still have the Fourier transform for $X$ and are able to perform numerical analysis.
\end{remark}

 Now we define the \emph{American payer swaption} as an option to enter at any time $T$ between $T_0$ and $T_n$ into an interest rate swap, paying the fixed leg at
a pre-determined rate $K$ and receiving the floating leg. In the Bermudan version there is finite number of dates when the holder can enter into the swap. We will formulate the American payer swaption
pricing problem as an optimal stopping problem.
For this we first note that the value of the swap at time $T\in[T_0,T_n]$ is
\begin{align} \label{payoff-4} \hs{0pc}
 \Pi^{swap}_T=\sum_{m=1}^{n} \Big(1-P(T,T_n)-(T_m\m T)KP(T,T_m)-\Delta K\sum_{j=m+1}^{n}P(T,T_j)\Big)1_{T_{m-1}\le T<T_m}
  \end{align}
where we take into account the accrual rate between $T$ and next payment date $T_m$ of the swap.
According to the definition of the state-price density, the price of the American swaption at time $T_0$ then can be expressed as the value function of the optimal stopping problem
\begin{align} \label{ost-0} \hs{6pc}
V^{A}(T_0,x)=\frac{1}{\zeta_{T_0}}\sup_{T_0\le\tau\le T_n}\EE_{T_0,x} \left[\zeta_{\tau} (\Pi^{swap}_{\tau})^+\right]
 \end{align}
where the supremum is taken over all stopping times $\tau$ with respect to $X$. In this paper we exploit a Markovian approach so that we introduce the following extension of \eqref{ost-0}
\begin{align} \label{ost-1} \hs{6pc}
V^{A}(t,x)=\frac{1}{\zeta_t}\sup_{t\le\tau\le T_n}\EE_{t,x} \left[\zeta_{\tau} (\Pi^{swap}_{\tau})^+\right]
 \end{align}
for $(t,x)\in [T_0,T_n]\times(0,\infty)$. Once \eqref{ost-1} is determined, one can compute the price $V^A(t,x)$ at time $t\in[0,T_0)$ as
\begin{align} \label{ost-2} \hs{6pc}
V^A(t,x)=\frac{1}{\zeta_t}\EE_{t,x} \left[\zeta_{T_0}V^A(T_0, X_{T_0})\right]
 \end{align}
 using the known distribution of $X_{T_0}$.
\vs{6pt}

3. Now using \eqref{cond-zeta} and \eqref{payoff-4} let us calculate the payoff in the optimal stopping problem \eqref{ost-1} when $\tau\in[T_{m-1},T_m)$ for every $m=1,...,n$
 \begin{align} \label{payoff-5} \hs{1pc}
 \zeta_{\tau}(\Pi^{swap}_{\tau})^+=&\Big[\zeta_{\tau}-\EE_{\tau}[\zeta_{T_n}]-(T_m\m \tau)K \EE_{\tau}[\zeta_{T_m}]
 -\Delta K\sum_{j=m+1}^{n}\EE_{\tau}[\zeta_{T_j}]\Big]^+\\
  =&\Big[\e^{-\int_0^{\tau} \alpha(s)ds}(1\p X_{\tau})
  -\e^{-\int_0^{T_n} \alpha(s)ds}(1\p\theta\p \e^{-\kappa(T_n-\tau)}(X_{\tau} \m\theta))\nonumber\\
 &-(T_m \m \tau)K \e^{-\int_0^{T_m} \alpha(s)ds}(1\p\theta\p\e^{-\kappa(T_m -\tau)}(X_{\tau} \m\theta))\nonumber\\
&-\Delta K\sum_{j=m+1}^{n}\e^{-\int_0^{T_j} \alpha(s)ds}\big(1\p\theta\p\e^{-\kappa(T_j -\tau)}(X_{\tau} \m\theta)\big)\Big]^+\nonumber\\
 =&\Big[G^1_m (\tau)X_{\tau}+G^2_m(\tau)\Big]^+\nonumber
   \end{align}
 where the functions $G^1_m$ and $G^2_m$ are given on intervals $[T_{m-1},T_m)$ by
 \begin{align} \label{payoff-A} \hs{3pc}
 G^1_m (t) =\;&\e^{-\int_0^t \alpha(s)ds}-c_n \e^{-\kappa(T_n -t)}-c_m(T_m \m t)K \e^{-\kappa(T_m -t)}
 -\Delta K\sum_{j=m+1}^{n}c_j \e^{-\kappa(T_j -t)}\\
 \label{payoff-B}
 G^2_m (t) =\;&\e^{-\int_0^t \alpha(s)ds}-c_n (1+\theta-\theta \e^{-\kappa (T_n -t)})
 -c_m (T_m \m t)K (1+\theta-\theta \e^{-\kappa (T_m -t)})\\
 &-\Delta K\sum_{j=m+1}^{n}c_j (1+\theta-\theta \e^{-\kappa (T_j -t)})\nonumber\\
 =\;&\theta(\wh{G}^1_m(t)-G^1_m (t))+\wh{G}^1_m(t)\nonumber
   \end{align}
where we define the coefficients $c_i:=\exp(-\int_0^{T_i} \alpha(s)ds)$, $i=1,...,n$, and $\wh{G}^1_m$ are given as $G^1_m$ in \eqref{payoff-A} with $\kappa=0$
 \begin{align} \label{payoff-whA} \hs{3pc}
 \wh{G}^1_m (t) =\e^{-\int_0^t \alpha(s)ds}-c_n-c_m (T_m \m t)K -\Delta K\sum_{j=m+1}^{n}c_j .
 \end{align}
Therefore we can formulate the following optimal stopping problem
\begin{align} \label{ost-3} \hs{6pc}
V(t,x)=\sup_{t\le\tau\le T_n}\EE_{t,x} \left[G(\tau,X_\tau)^+\right]
 \end{align}
for $(t,x)\in [T_0,T_n]\times(0,\infty)$ and the function $G$ is given by
 \begin{align} \label{payoff-6} \hs{3pc}
G(t,x)=\sum_{m=1}^{n} \big(G^1_m(t)x+G^2_m(t)\big)1_{T_{m-1}\le t<T_m}=G^1(t)x+G^2(t)
 \end{align}
 for $(t,x)\in [T_0,T_n]\times(0,\infty)$ and where the functions $G^1$ and $G^2$ are given piecewisely on intervals
 $[T_{m-1},T_m)$ by $G^1_m$ and $G^2_m$ as
 \begin{align} \label{payoff-AA} \hs{3pc}
 G^1(t) =\sum_{m=1}^{n} G^1_m(t)1_{T_{m-1}\le t<T_m} \quad\text{and}\quad G^2(t) =\sum_{m=1}^{n} G^2_m(t)1_{T_{m-1}\le t<T_m}
   \end{align}
 for $t\in[T_0,T_n]$.
Using \eqref{ost-1}, \eqref{payoff-5} and \eqref{ost-3} we obtain
\begin{align} \label{ost-4} \hs{6pc}
V^{A}(t,x)=V(t,x)/\zeta_t=\e^{\int_0^t \alpha(s)ds}\,V(t,x)/(1\p X_t)
 \end{align}
 for $(t,x)\in [T_0,T_n]\times(0,\infty)$ so that we now focus on the problem \eqref{ost-4}.

 It is important to note that $G(T_n,x)=0$ for all $x>0$ and hence it is not optimal to stop
 in the set where $G\le 0$ since with positive probability we can enter later into the set where $G>0$. This observation allows us to simplify \eqref{ost-4} by removing the positive part
and formulate the equivalent problem
\begin{align} \label{ost-5} \hs{6pc}
V(t,x)=\sup_{t\le\tau\le T_n}\EE_{t,x} \left[G(\tau,X_\tau)\right]
 \end{align}
for $(t,x)\in [T_0,T_n]\times(0,\infty)$.

\vs{6pt}

4. Now we turn to the \emph{American receiver swaption} which is the option to enter at any time $T$ between $T_0$ and $T_n$ into an interest rate swap, receiving the fixed leg at
a pre-determined rate $K$ and paying the floating leg. The value of the swap, from the
perspective of the fixed-rate receiver, has the same absolute value as in \eqref{payoff-1} but the opposite sign. Therefore by doing similar manipulations as in paragraph 3 above
we are delivered the following optimal stopping problem
\begin{align} \label{ost-6} \hs{6pc}
\wt{V}(t,x)=\inf_{t\le\tau\le T_n}\EE_{t,x} \left[G(\tau,X_\tau)\right]
 \end{align}
 and the price of American receiver swaption is
 \begin{align} \label{ost-7} \hs{6pc}
\wt{V}^{A}(t,x)=-\e^{\int_0^t \alpha(s)ds}\wt{V}(t,x)/(1\p X_t)
 \end{align}
 for $(t,x)\in [T_0,T_n]\times(0,\infty)$.
 Since the \eqref{ost-5} is a minimization problem and $G(T_n,x)=0$ for all $x>0$ it is obvious that one should not stop in the set where $G$ is positive. Both problems
 \eqref{ost-5} and \eqref{ost-6} have the same gain function, however the former is a maximization problem and the latter is a minimization problem. We will analyze  \eqref{ost-5} in the next section and will discuss briefly the solution to
 \eqref{ost-6} in Section~\ref{secrec}.

 \section{Free-boundary problem for fixed-rate payer}\label{secpay}

 In this section we will reduce the problem \eqref{ost-5} to a free-boundary problem and the latter will be tackled using local time-space calculus (\cite{Pe-1}).
 First using that the gain function $G(t,x)$ is continuous and standard arguments (see e.g. Corollary 2.9 (Finite horizon) with Remark 2.10 in \cite{PS}) we have that continuation and stopping sets read
\begin{align} \label{C} \hs{5pc}
&C^*= \{\, (t,x)\in[T_0,T_n)\! \times\! (0,\infty):V(t,x)>G(t,x)\, \} \\[3pt]
 \label{D}&D^*= \{\, (t,x)\in[T_0,T_n)\! \times\! (0,\infty):V(t,x)=G(t,x)\, \}
 \end{align}
and the optimal stopping time in \eqref{ost-5} is given by
\begin{align} \label{OST} \hs{5pc}
\tau^*=\inf\ \{\ t\leq s\leq T_n:(s,X^x_{s})\in D^*\ \}.
 \end{align}

In view of the bounds \eqref{srbounds} it follows that the model implied forward and swap rates are essentially bounded from below and above by $\sup_{t>0} \alpha(t)-\kappa\theta$ and $\inf_{t>0} \alpha(t)+\kappa$, respectively. More precisely, these bounds are exact when $\alpha(t)\equiv\alpha$ is constant and close to exact when $\sup_{t>0} \alpha(t)$ is close to $\inf_{t>0} \alpha(t)$. Hence a payer (receiver) swaption with strike rate above $\inf_{t>0} \alpha(t)+\kappa$ (below $\sup_{t>0} \alpha(t)-\kappa\theta$) would trivially have zero value. We thus henceforth assume that the strike rate $K$ lies in the range
\begin{align} \label{assK} \hs{5pc}
K\begin{cases} \le \inf_{t>0} \alpha(t)+\kappa, &\text{for a payer swaption}\\
\ge \sup_{t>0} \alpha(t)-\kappa\theta &\text{for a receiver swaption.}
\end{cases}
 \end{align}
From 1$(iii)$ below we will see that the function $G^1$, the leading term in \eqref{payoff-6}, is positive on $[T_{0},T_n)$ under this condition.
If it does not hold and, say $\kappa\p \alpha(T_n)<K$, then $G^1 <0$ at least in some neighborhood of $T_n$ and thus $G^2<0$ is negative as well so that $G<0$ and it is not optimal to exercise the swap at that period of time for any value of factor process $X$. Moreover, the exercise set will have a very complicated structure and the problem \eqref{ost-5} has to be tackled case by case.
\vs{6pt}

1. Below we provide some properties of the functions $G^1$ and $G^2$.
\vs{+4pt}

$(i)$ It is obvious that $G^1(T_n)=G^2(T_n)=0$. From \eqref{payoff-A}-\eqref{payoff-whA} and the fact that $G^1_m\ge\wh{G}^1_m$ everywhere, it follows that $G^1(t)>G^2(t)$ for all $t\in [T_{0},T_n)$.
\vs{+2pt}

$(ii)$ We show that $G^1$ and $G^2$ are continuous on $[T_{0},T_n)$, however their derivatives, in general, discontinuous at payment dates $T_m$, $m=1,...,n-1$.
From \eqref{payoff-A}-\eqref{payoff-B} we see that $G^1_m$ and $G^2_m$ are smooth on $[T_{m-1},T_m)$ for $m=1,...,n$. We then observe that functions $G^1$ and $G^2$ are continuous at payment dates $T_m$, $m=1,...,n-1$
\begin{align} \label{G-2} \hs{1pc}
G^1(T_m -)&=G^1_m(T_m-)=c_m-c_n \e^{-\kappa (T_n -T_m)}-\Delta K\sum_{j=m+1}^{n}c_j \e^{-\kappa(T_j -T_m)}\\
&=G^1_{m+1} (T_m)=G^1(T_m +)\nonumber\\
G^2(T_m -)&=G^2_m(T_m-)=\theta(\wh{G}^1_m(T_m-)-G^1_m (T_m-))+\wh{G}^1_m(T_m-)\nonumber\\
&=\theta(\wh{G}^1_{m+1}(T_m)-G^1_{m+1} (T_m))+\wh{G}^1_{m+1}(T_m)=G^2_{m+1} (T_m)=G^2(T_m +)\nonumber
\end{align}
at payment dates $T_m$, $m=1,...,n-1$. Now by straightforward calculations of the derivatives of $G^1_m$ and $G^2_m$ we have
\begin{align} \label{G-3} \hs{1pc}
(G^1_m)'(t)=&\kappa\big[\m c_n \e^{-\kappa (T_n -t)}\m c_m(T_m \m t)K \e^{-\kappa(T_m -t)}\m\Delta K\sum_{j=m+1}^{n}c_j \e^{-\kappa(T_j -t)}\big]\\
&-\alpha(t)\e^{-\int_0^{t} \alpha(s)ds}+c_m Ke^{-\kappa(T_m -t)}\nonumber\\
=&\kappa\big[G^1_m (t)-\e^{-\int_0^{t} \alpha(s)ds}\big]-\alpha(t)\e^{-\int_0^{t} \alpha(s)ds}+c_m K\e^{-\kappa(T_m -t)}\nonumber\\
\label{G-4}
(G^2_m)' (t)=&\theta\left((\wh{G}^1_m)'(t)-(G^1_m)' (t)\right)+(\wh{G}^1_m)'(t)=(1\p\theta)(\wh{G}^1_m)'(t)-\theta (G^1_m)'(t)
\end{align}
for $t\in[T_{m-1},T_m)$ and $m=1,...,n$. Then it follows from \eqref{payoff-A}-\eqref{payoff-B} that the derivatives of $G^1$ and $G^2$ generally
(except from specifically chosen set of parameters) are not continuous at $T_m$
\begin{align} \label{G-5}  \hs{3pc}
(G^1)'(T_m -)-&(G^1)'(T_m +)=c_m K-c_{m+1}Ke^{-\kappa\Delta}>0\\
\label{G-6}
(G^2)'(T_m -)-&(G^2)'(T_m +)=K-(1\p\theta)K+\theta Ke^{-\kappa\Delta}<0
\end{align}
for $m=1,...,n-1$.
\vs{+2pt}

$(iii)$ Here we show that, due to assumption~\eqref{assK}, the function $G^1$ is positive on $[T_{0},T_n)$.
From \eqref{G-3} we have that $G^1$ is increasing at point $t\in[T_{m-1},T_m)$ if and only if
\begin{align} \label{G-7} \hs{3pc}
G^1(t)>\frac{1}{\kappa}\left[\e^{-\int_0^{t} \alpha(s)ds}\left(\kappa\p \alpha(t)\right)-c_m K e^{-\kappa(T_m -t)}\right]:=\pi(t)
\end{align}
and thus we need to compare $G^1$ itself with the function $\pi$ which is right-continuous with jumps at payment dates $T_m$, $m=1,...,n-1$. The function $\pi$ is positive
\begin{align} \label{G-8} \hs{3pc}
\pi(t)=\frac{\e^{-\int_0^{t} \alpha(s)ds}}{\kappa}\left[\kappa\p \alpha(t)-K \e^{-\int_t^{T_m} (\kappa+\alpha(s))ds}\right]>0
\end{align}
for $t\in[T_{m-1},T_m)$, $m=1,...,n$, by using~\eqref{assK}.
We then note that $G^1 (T_n)=0$ and $\pi(T_n)=c_n(\kappa+ \alpha(T_n)- K)/\kappa>0$.
Therefore we have that $G^1<\pi$ near $T_n$ and thus $G^1$ is decreasing and positive there.
Then the fact that $G^1$ is positive on $[T_0, T_n]$ can be shown by going backward from $T_n$ and using two observations:
$a)$ when $G^1<\pi$ the function $G^1$ goes far away from 0 and $b)$ when $G^1>\pi$ it dominates positive function $\pi$.
\vs{2pt}

$(iv)$ Now we consider the limits of $G^1$ and $G^2$ near $t=T_n$. By L'Hospital's rule we figure out that
\begin{align} \label{G1-limit}  \hs{7pc}
&\lim_{t\uparrow T_n}\frac{G^1(t)}{T_n -t}=c_n\left(\alpha(T_n)\p\kappa\m K\right)>0\\
\label{G2-limit}
&\lim_{t\uparrow T_n}\frac{G^2(t)}{T_n -t}=c_n\left(-\theta \kappa\p \alpha(T_n)\m K\right).
\end{align}

3. In order to get some insights into the structure of stopping region $D^*$ we first need to calculate the function
\begin{align} \label{H-0} \hs{7pc}
H(t,x)=(G_t \p \L_X G)(t,x)
\end{align}
for $(t,x)\in[T_0,T_n)\times(0,\infty)$ where
$\L_X =\kappa (\theta-x)d/dx+(\sigma^2/2)\, x\, d^2/dx^2$ is the infinitesimal generator of $X$.
The function $H$ has the meaning of the
instantaneous benefit of waiting to exercise. By straightforward calculations we have that
\begin{align} \label{H-1} \hs{3pc}
H(t,x)=\sum_{m=1}^{n} \big(H^1_m(t)x+H^2_m(t)\big)1_{T_{m-1}\le t<T_m}=H^1(t)x+H^2(t)
\end{align}
for $(t,x)\in[T_{0},T_n)\times(0,\infty)$ where
\begin{align} \label{H-2} \hs{3pc}
 H^1_m (t) =& -\left(\kappa\p \alpha(t)\right)\e^{-\int_0^{t} \alpha(s)ds}+c_m K e^{-\kappa(T_m -t)}=-\kappa\pi(t)\\
 \label{H-3}
 H^2_m (t) =& \;-\theta H^1_m (t)+(1\p\theta)\left(c_m K \m\alpha(t)\e^{-\int_0^{t} \alpha(s)ds}\right)
   \end{align}
for $t\in[T_{m-1},T_m)$, $m=1,...,n$, and
\begin{align} \label{H-4} \hs{3pc}
 H^1(t) =\sum_{m=1}^{n}H^1_m(t)1_{T_{m-1}\le t<T_m}\quad\text{and}\quad H^2(t) =\sum_{m=1}^{n}H^2_m(t)1_{T_{m-1}\le t<T_m}
   \end{align}
for $t\in[T_{0},T_n)$.
Therefore we have that
$t\mapsto H(t,x)$ is right-continuous with jumps at $T_m$, $m=1,....,n\m1$,
for each $x>0$ fixed. We also observe from \eqref{H-2} that the function $H^1<0$ on $[T_{0},T_n)$ as we proved above that the function $\pi>0$ on $[T_{0},T_n)$.

Since the function $G$ is not $C^1$ at payment dates $T_m$, $m=1,....,n\m1$ with respect to time variable, we have to use It\^{o}-Tanaka formula to get
\begin{align} \label{Ito-1} \hs{3pc}
 \EE \left[G(\tau,X^{t,x}_\tau)\right]=\;&G(t,x)+\EE\left[ \int_t^\tau H(s,X^{t.x}_s)1_{\{s\neq T_m, m=1,..,n-1\}}ds\right]\\
 =\;&G(t,x)+\EE\left[ \int_t^\tau H(s,X^{t.x}_s)ds\right]\nonumber
 \end{align}
for $(t,x)\in[T_0,T_n)\times(0,\infty)$ where the integral term with respect to the local time is not present since the underlying process, time, is of bounded variation,
and in the last integral we omitted the indicator of Lebesgue-measure null set.
It is obvious that the expression \eqref{Ito-1} indicates that the set $\{(t,x)\in [T_{0},T_n)\times(0,\infty): H(t,x)>0\}$ belongs to continuation set $C^*$
(for this one can make use of the first exit time from a
sufficiently small time-space ball centred at the point where $H>0$).

\vs{6pt}

4.  Next we show up-connectedness of the stopping set $D^*$.
 For this, let us take two points $(t,x)$ and $(t,y)$ such that $t\in[T_0,T_n)$ and $y>x>0$, then let us denote by $\tau=\tau^* (t,y)$ the optimal stopping time for $V(t,y)$ so that
 using \eqref{Ito-1} we have
 \begin{align} \label{b-0} \hs{3pc}
 V(t,x)-V(t,y)\ge\;& \EE\left[ G(\tau,X^{t,x}_\tau)\right]-\EE \left[G(\tau,X^{t,y}_\tau)\right]\\
 =\;&G(t,x)-G(t,y)+\EE \left[ \int_t^\tau \big(H(s,X^{t,x}_s)-H(s,X^{t,y}_s)\big)ds\right]\nonumber\\
 =\;&G(t,x)-G(t,y)+\EE  \left[ \int_t^\tau \
  H^1(s)(X^{t,x}_s \m X^{t,y}_s)ds\right]\nonumber\\
 \ge\;&G(t,x)-G(t,y)\nonumber
 \end{align}
 where in the last inequality we used facts that the function $H^1$ is negative and $X^{t,y}_\cdot \ge X^{t,x}_\cdot$  $\PP$-a.s. by the comparison principle for SDEs.
 Now if we take $(t,x)\in D^*$, i.e. $V(t,x)=G(t,x)$,  we have that $V(t,y)=G(t,y)$ and thus $(t,y)\in D^*$.
 Therefore there exists a function $b:[T_0,T_n)\rightarrow (0,\infty)$ such that
 \begin{align} \label{b-2} \hs{5pc}
D^*= \{\, (t,x)\in[T_0,T_n)\! \times\! (0,\infty):x\ge b(t)\, \}.
 \end{align}

A direct examination of functions $G$ and $H$ in \eqref{payoff-6} and \eqref{H-1} imply that there exist
real-valued curves $g$ and $h$ on $[T_0,T_n]$ defined as
\begin{align} \label{b-1} \hs{5pc}
G(t,g(t)) = 0 \quad\text{and}\quad H(t,h(t)) = 0
   \end{align}
for $t\in [T_{0},T_n)$ (see Figure 1) such that
$G(t,x)>0$ for $x>g(t)$ and $G(t,x)<0$ for $x<g(t)$, $H(t,x)>0$ for $x<h(t)$
and $H(t,x)<0$ for $x>h(t)$ when
$t\in[T_0,T_n)$ is given and fixed.
It is not optimal to stop when $G<0$ or $H>0$ so that we have $b>g\vee 0$ and $b>h\vee 0$ on $[T_{0},T_n)$ as $X_t$ is always positive.

It is clear that if $x>h(t)\vee 0$ and $t<T_n$ is sufficiently close to $T_n$ then it is optimal to stop
immediately (since the profit obtained from being below $h$ cannot compensate the cost of getting
there due to the lack of remaining time). This shows that $b(T_n -)=h(T_n -)\vee 0$ where
 \begin{align} \label{h-1} \hs{5pc}
h(T_n -)=-\frac{H^2 (T_n-)}{H^1 (T_n-)}=\frac{\theta \kappa\m \alpha(T_n)\p K}{\alpha(T_n)\p\kappa\m K}
\end{align}
as is easily seen from \eqref{H-2}-\eqref{H-3}.
We also notice that
 \begin{align} \label{g-1} \hs{5pc}
g(T_n -)=-\frac{G^2 (T_n-)}{G^1 (T_n-)}=\frac{\theta \kappa\m \alpha(T_n)\p K}{\alpha(T_n)\p\kappa\m K}=h(T_n-)
\end{align}
by \eqref{G1-limit}-\eqref{G2-limit}.

\vs{6pt}

\begin{figure}[t]
\begin{center}
\includegraphics[scale=0.7]{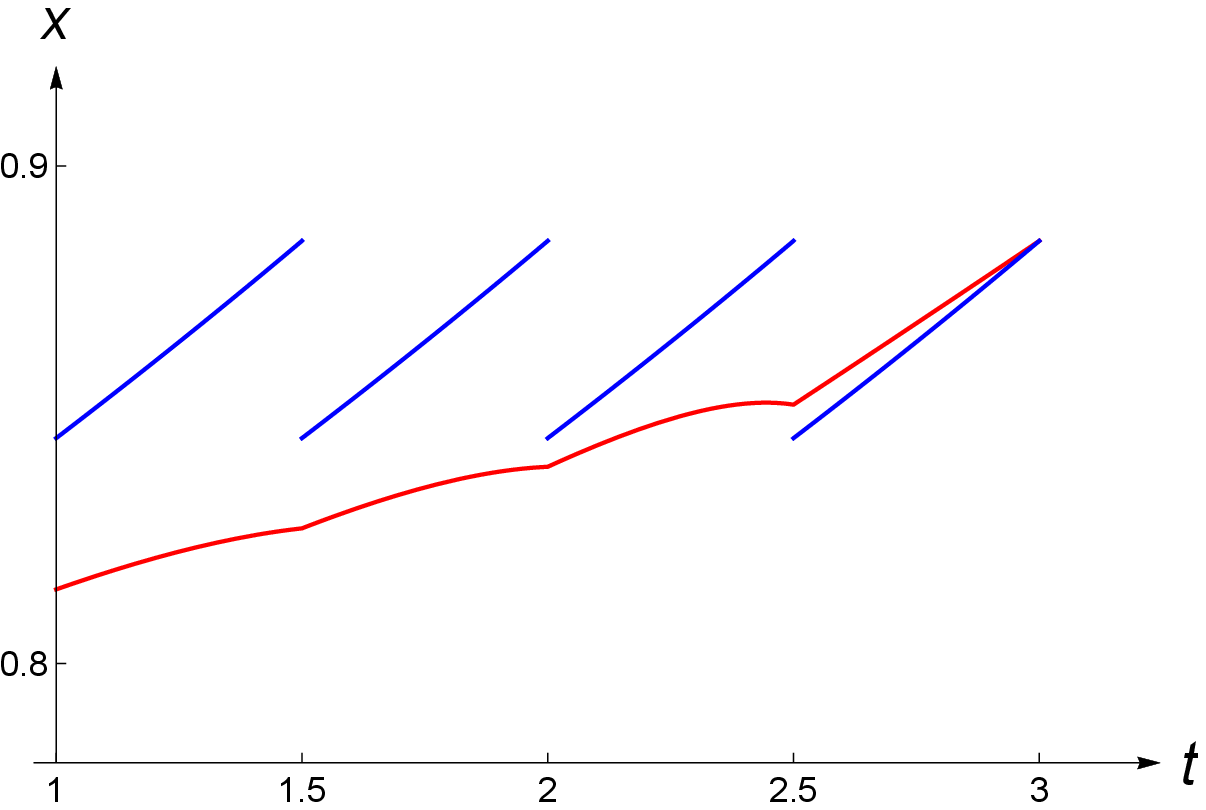}
\end{center}

{\par \leftskip=1.6cm \rightskip=1.6cm \small \ni \vs{-10pt}

\textbf{Figure 1.} A computer drawing of the functions $g$ (red line) and $h$ (blue line) defined in \eqref{b-1}.
The parameter set is $T_0=1$ year, $\Delta=0.5$ year, $n=4$, $K=0.05$, $\theta=2.55$, $\kappa=0.03$,  $\alpha\equiv \theta\kappa=0.0765$. Coefficient $\sigma$ is time-dependent function and calibrated from European swaption prices.

\par} \vs{10pt}

\end{figure}

4. Standard Markovian arguments lead to the following free-boundary problem (for the value
function $V=V(t,x)$ and the optimal stopping boundary $b=b(t)$ to be determined):
\begin{align} \label{PDE} \hs{5pc}
&V_t \p\L_X V=0 &\hs{-30pt}\text{in}\;  C^*\\
\label{IS}&V(t,b(t))=G(t,b(t)) &\hs{-30pt}\text{for}\; t\in[T_0,T_n)\\
\label{SF}&V_x (t,b(t))=G_x(t,b(t)) &\hs{-30pt}\text{for}\; t\in[T_0,T_n) \\
\label{FBP1}&V(t,x)>G(t,x) &\hs{-30pt}\text{in}\; C^*\\
\label{FBP2}&V(t,x)=G(t,x) &\hs{-30pt}\text{in}\; D^*
\end{align}
where the continuation set $C^*$ and the stopping set $D^*$ are given by
\begin{align} \label{C-1} \hs{5pc}
&C^*= \{\, (t,x)\in[T_0,T_n)\! \times\! (0,\infty):x<b(t)\, \} \\[3pt]
 \label{D-1}&D^*= \{\, (t,x)\in[T_0,T_n)\! \times\! (0,\infty):x\ge b(t)\, \}.
 \end{align}
It can be shown that this free-boundary problem has a unique solution $V$ and $b$ which coincide with the
value function \eqref{ost-5} and the optimal stopping boundary respectively (cf. \cite{PS}).

Completed details of the analysis above go beyond our goals in this paper and for this reason
will be omitted. It should be noted however that one of the main issues which makes this
analysis quite complicated (in comparison with e.g. the American put option problem)
is that $b$ seems not monotone function of time. This fact is supported by numerical analysis (see Figure 2 near $t=0$).
The standard probabilistic intuition and proof of monotonicity requires that $H_t <0$ however straightforward differentiation in \eqref{H-1}
shows that it is not true. Proof of the continuity of the free boundary without having its
monotonicity is open and challenging problem, which can help to tackle other optimal stopping problems.
In the next section we will derive simpler equations
which characterize $V$ and $b$ uniquely and can be used for financial analysis of American swaptions.

\vs{6pt}

5. We now provide the early exercise premium representation formula for the value function $V$ which decomposes it into the sum of the expected payoff
with exercise at $T_n$ (which is zero) and early exercise premium which depends on the boundary $b$. The optimal stopping boundary $b$ will be obtained as the unique solution to the nonlinear integral equation of Volterra type.
We denote the following function
\begin{align} \label{L} \hs{5pc}
L(t,u,x,z)=-\EE_{t,x} \left[H(u,X_u) I(X_u \ge z)\right]
 \end{align}
 for $u\ge t \ge 0$ and $x,z>0$. If the probability density function $p(\wh{x};u,x,t)$ of  $X_u$ under $\EE_{t,x}$ is known (when $\sigma(t)\equiv \sigma$ so that $X$ has non-central chi-squared distribution, see \cite{CIR}), then $L$ can be computed as
 as follows
\begin{align} \label{LL} \hs{5pc}
L(t,u,x,z)=-\int_z^\infty H(u,\wh{x})p(\wh{x};u,x,t)d\wh{x}.
 \end{align}
 by univariate numerical integration.

 Otherwise, in the case of time-dependent $\sigma$ we can obtain the Fourier transform $q(z;u,x,t)$ for $X_u$
  \begin{align} \label{Fourier} \hs{5pc}
q(w;u,x,t)=\EE_{t,x}\left[ e^{wX_u} \right]=e^{\varphi(w,u,t)+x\phi(w,u,t)}
 \end{align}
 for $w\in \mathbb{C}$, $u\ge t\ge 0$ and $x>0$, where $\varphi$ and $\phi$ are obtained by solving corresponding Ricatti equations (see e.g. \cite{F-2})
   \begin{align} \label{Fourier-2} \hs{5pc}
&\varphi(w,u,t)=\kappa\, \theta \int_t^u \frac{e^{-\kappa(v-t)}w}{1-\frac{w}{2}\int_t^v e^{-\kappa(v-s)}\sigma^2(s)ds}dv \\
&\phi(w,u,t)=\frac{e^{-\kappa(u-t)}w}{1-\frac{w}{2}\int_t^u e^{-\kappa(u-s)}\sigma^2(s)ds}.
 \end{align}
  We can then recover the probability density function $p$ of $X_u$ as follows
  \begin{align} \label{pdf} \hs{5pc}
p(\wh{x};u,x,t)=\frac{1}{2\pi}\int_{\R} e^{-iw\wh{x}}q(iw;u,x,t)dw
 \end{align}
for $\wt{x},x\ge 0$ and $u\ge t\ge 0$. Moreover, using Theorem 4 in \cite{FLT} we have that
    \begin{align} \label{th4} \hs{5pc}
\EE_{t,x}\left[(X_u-z)^+\right]=\frac{1}{\pi}\int_0^\infty \mathbf{Re}\left[\frac{q(\mu+i\lambda;u,x,t)}{e^{(\mu+i\lambda)z}(\mu+i\lambda)^2}\right]d\lambda
 \end{align}
  for $u\ge 0$ and $x,z>0$, where $\mu>0$ is chosen such that $q(\mu;u,x,t)<\infty$. Hence, we can compute $L$ in the following way
  \begin{align} \label{LLL}
-L(t,u,x,z)=&\EE_{t,x} \left[H(u,X_u) I(X_u \ge z)\right]\\
=&H^1(u)\EE_{t,x} \left[X_u I(X_u \ge z)\right]+H^2(u)\PP_{t,x}(X_u \ge z)\notag\\
=&H^1(u)\EE_{t,x} \left[(X_u-z)^+ \right]+(H^2(u)+H^1(u)z)\PP_{t,x}(X_u \ge z)\notag\\
=&H^1(u)\frac{1}{\pi}\int_0^\infty \mathbf{Re}\left[\frac{q(\mu+i\lambda;u,x,t)}{e^{(\mu+i\lambda)z}(\mu+i\lambda)^2}\right]d\lambda\notag\\
&+(H^2(u)+H^1(u)z)\left(\frac{1}{2}+\frac{1}{\pi}\int_0^\infty \frac{\mathbf{Im}[e^{-i\lambda z}q(i\lambda;u,x,t)]}{\lambda}d\lambda\right)\notag
 \end{align}
where in the last equality we used the formula for the cumulative distribution function via Fourier transform. Thus the computation of $L$ in this case requires univariate integration as in \eqref{LL}.
\vs{2pt}

 The main result of this section may now be stated as follows.
\begin{theorem}\label{th:3.1}
The value function $V$ of \eqref{ost-5} has the following representation
\begin{align}\label{th-1} \hs{5pc}
V(t,x)=\int_t^{T_n}L(t,u,x,b(u))du
\end{align}
for $t\in[T_0,T_n)$ and $x\in (0,\infty)$. The optimal stopping boundary $b$ in \eqref{ost-5} (see Figure 2) can be characterized as the unique solution to the nonlinear integral equation
\begin{align}\label{th-2} \hs{5pc}
G(t,b(t))=\int_t^{T_n} L(t,u,b(t),b(u))du
\end{align}
for $t\in[T_0,T_n)$ in the class of continuous functions $t\mapsto b(t)$
with $b(T_n)=\frac{\theta \kappa- \alpha(T_n)+ K}{\alpha(T_n)+\kappa- K}\vee 0$.
\end{theorem}

\begin{figure}[t]
\begin{center}
\includegraphics[scale=0.7]{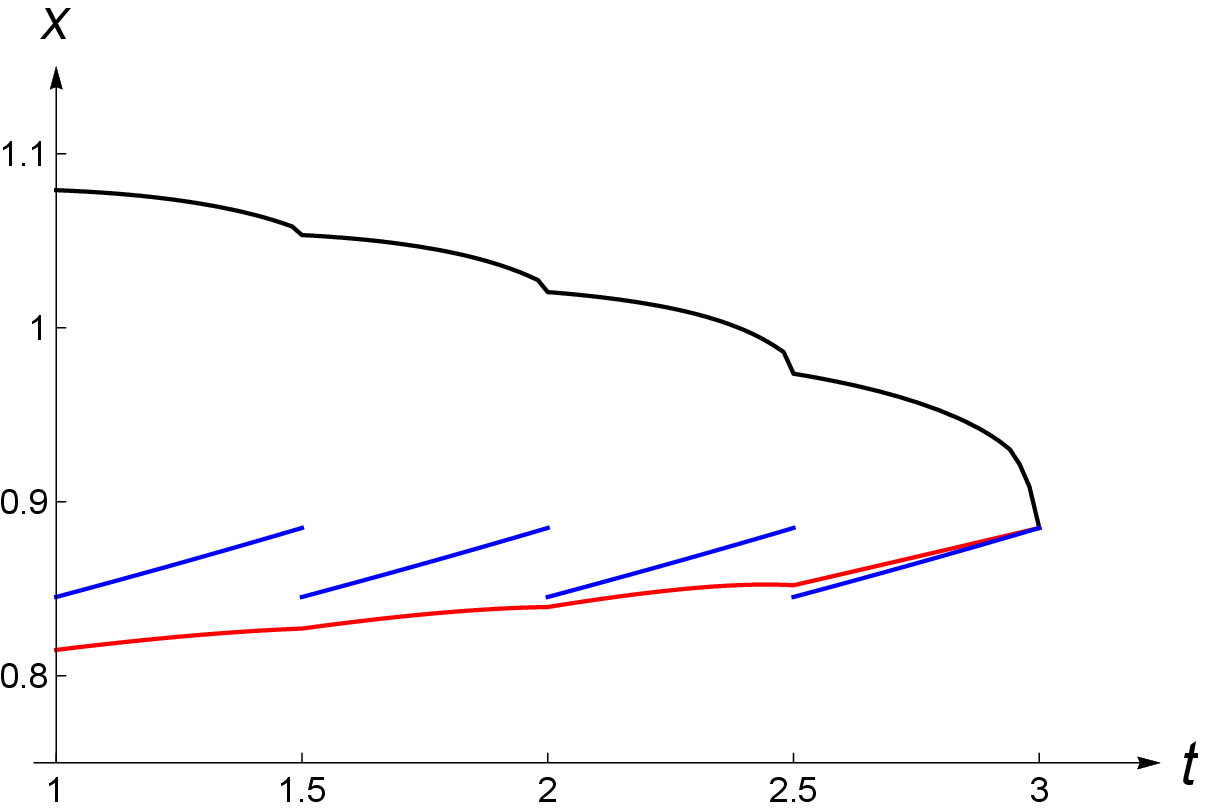}
\end{center}

{\par \leftskip=1.6cm \rightskip=1.6cm \small \ni \vs{-10pt}

\textbf{Figure 2.} A computer drawing of the optimal stopping boundary
$t\mapsto b(t)$ for the problem \eqref{ost-5}.
The parameter set is $T_0=1$ year, $\Delta=0.5$ year, $n=4$, $K=0.05$, $\theta=2.55$, $\kappa=0.03$,  $\alpha\equiv \theta\kappa=0.0765$. Coefficient $\sigma$ is time-dependent function and calibrated from European swaption prices.

\par} \vs{10pt}

\end{figure}
\begin{proof}

$(A)$ By applying the local time-space formula on curves \cite{Pe-1} for
$V(s,X_s)$ we have that
\begin{align} \label{th-a} \hs{1pc}
V(s&,X_s)\\
=\;&V(t,x)+M_s\nonumber\\
 &+ \int_t^{s}
(V_t \p\L_X V)(u,X_u)
 I(X_u \neq b(u))du\nonumber\\
 &+\frac{1}{2}\int_t^{s}
\big(V_x (u,X_u +)\m V_x (u,X_u -)\big)I\big(X_u=b(u)\big)d\ell^{b}_u(X)\nonumber\\
  =\;&V(t,x)+M_s + \int_t^{s}
(G_t \p\L_X G)(u,X_u)I(X_u \ge b(u))du\nonumber\\
  =\;&V(t,x)+M_s +\int_t^{s}
H(u,X_u) I(X_u \ge b(u))du\nonumber
  \end{align}
where we used \eqref{PDE}, the definition of $H$  \eqref{H-0}, the smooth-fit condition \eqref{SF} and where $M=(M_u)_{u\ge t}$ is the martingale term,  $(\ell^{b}_u(X))_{u\ge t}$ is the local time process of $X$ spending at the boundary $b$. Now
upon letting $s=T_n$, taking the expectation $\EE_{t,x}$,  using the optional sampling theorem for $M$, rearranging terms and noting that $V(T_n,x)=G(T_n,x)=0$ for all $x>0$, we get \eqref{th-1}.
The integral equation \eqref{th-2} is obtained by inserting $x=b(t)$ into \eqref{th-1} and using \eqref{IS}.

\vs{6pt}

$(B)$ Now we show that $b$ is the unique solution to the equation \eqref{th-2} in the class of continuous functions $t\mapsto b(t)$.   The proof is divided in few steps and it is based on arguments originally derived in \cite{Pe-4}.

\vs{6pt}

$(B.1)$ Let $c:[T_0,T_n]\rightarrow \R$  be a solution to the equation
\eqref{th-2} such that $c$ is continuous. We will show that these $c$ must be equal to the optimal stopping boundary $b$.
Now let us consider the function $U^{c}:[T_0,T_n)\times(0,\infty)\rightarrow \R$ defined as follows
\begin{align}\label{th-b} \hs{3pc}
U^{c}(t,x)=\;\int_t^{T_n} L(t,u,x,c(u))du
\end{align}
for $(t,x)\in[T_0,T_n]\times(0,\infty)$. Observe the fact that $c$ solves the equation
\eqref{th-2} means exactly that $U^{c}(t,c(t))=G(t,c(t))$ for all $t\in [T_0,T_n]$. We will moreover show that
$U^{c}(t,x)=G(t,x)$ for $x\in[c(t),\infty)$ with $t\in [T_0,T_n]$. This can be derived using martingale property as follows;
the Markov property of $X$ implies that
\begin{align}\label{th-5} \hs{3pc}
U^{c}(s,X_s)-\int_t^s H(u,X_u) I(X_u \ge c(u))du=\;U^{c}(t,x)+N_s
\end{align}
where $(N_s)_{t\le s\le T_n}$ is a martingale under $\PP_{t,x}$. On the other hand,
we know from \eqref{Ito-1}
\begin{align} \label{th-6} \hs{3pc}
G(s,X_s)=\;G(t,x)+\int_t^s H(u,X_u) du+M_s
 \end{align}
where $(M_s)_{t\le s\le T_n}$ is a continuous martingale under $\PP_{t,x}$.

For $x\in[c(t),\infty)$ with $t\in[T_0,T_n]$ given and fixed, consider the stopping time
 \begin{align} \label{th-7} \hs{3pc}
\sigma_{c}=\inf\ \{\ t\leq s\leq T_n:X_{s}\le c(s) \ \}
 \end{align}
 under $\PP_{t,x}$. Using that $U^{c}(t,c(t))=G(t,c(t))$ for all $t\in [T_0,T_n]$ and $U^{c}(T_n,x)=G(T_n,x)=0$ for all $x>0$, we see that
 $U^{c}(\sigma_{c},X_{\sigma_{c}})=G(\sigma_{c},X_{\sigma_{c}})$. Hence
 from \eqref{th-5}, \eqref{th-6} and  \eqref{th-7} using the optional sampling theorem we find:
 \begin{align}\label{th-8} \hs{3pc}
U^{c}(t,x)=\;&\EE_{t,x} \left[U^{c}(\sigma_{c},X_{\sigma_{c}})\right]-\EE_{t,x}\left[\int_t^{\sigma_{c}}H(u,XG_u) I(XG_u \ge c(u))du\right]\\
=\;&\EE_{t,x} \left[G(\sigma_{c},X_{\sigma_{c}})\right]-\EE_{t,x}\left[\int_t^{\sigma_{c}} H(u,X_u)du\right] =G(t,x)\nonumber
\end{align}
 since $X_u \in (c(t\p u),\infty)$ for all $u\in [0,\sigma_{c})$. This proves that $U^{c}(t,x)=G(t,x)$ for $x\in[c(t),\infty)$ with $t\in [T_0,T_n]$ as claimed.

 \vs{6pt}

 $(B.2)$ We show that $U^{c}(t,x)\le V(t,x)$ for all $(t,x)\in[T_0,T_n]\times(0,\infty)$.
 For this consider the stopping time
 \begin{align} \label{th-9} \hs{3pc}
\tau_{c}=\inf\ \{\ t\leq s\leq T_n:X_s\ge c(s)\ \}
 \end{align}
 under $\PP_{t,x}$ with $(t,x)\in[T_0,T_n]\times(0,\infty)$ given and fixed. The same arguments as those
 following \eqref{th-7} above show that $U^{c}(\tau_{c},X_{\tau_{c}})=G(\tau_{c},X_{\tau_{c}})$. Inserting $\tau_{c}$ instead of $s$ in \eqref{th-5} and using the optional sampling theorem, we get:
 \begin{align}\label{th-10} \hs{3pc}
U^{c}(t,x)=\EE_{t,x} \left[ U^{c}(\tau_{c},X_{\tau_{c}})\right]=
\EE_{t,x} \left[ G(\tau_{c},X_{\tau_{c}})\right]\le V(t,x)
\end{align}
 proving the claim.

 \vs{6pt}

  $(B.3)$ We show that $c\le b$ on $[T_0,T_n]$. For this, suppose that there exists $t\in[T_0,T_n)$ such that $b(t)< c(t)$
 and choose a point $x\in [c(t),\infty)$ and consider the stopping time
  \begin{align} \label{th-11} \hs{3pc}
\sigma=\inf\ \{\ t\leq s\leq T_n :b(s)\ge X_{s}\ \}
 \end{align}
 under $\PP_{t,x}$. Inserting $\sigma$ instead of $s$ in \eqref{th-a} and \eqref{th-5} and using the optional sampling theorem, we get:
 \begin{align} \label{th-12a} \hs{3pc}
&\EE_{t,x}  \left[ V(\sigma,X_\sigma)\right]=V(t,x) +\EE_{t,x} \left[\int_t^\sigma H(u,X_u) du\right]\\
\label{th-12b}&\EE_{t,x}  \left[U^{c}(\sigma,X_\sigma)\right]=U^{c}(t,x)+\EE_{t,x} \left[\int_t^{\sigma} H(u,X_u) I\big(X_u \ge c(u))\big)du\right].
\end{align}
Since $U^{c}\le V$ and $V(t,x)=U^{c}(t,x)=G(t,x)$ for
$x\in[c(t),\infty)$ with $t\in [T_0,T_n]$, it follows
from \eqref{th-12a} and \eqref{th-12b} that:
\begin{align} \label{th-13} \hs{3pc}
\EE_{t,x} \left[\int_t^{\sigma} H(u,X_u) I\big(X_u\le c(u)\big)du\right]\ge0.
\end{align}
Due to the fact that $H$ is strictly negative above $b$ we see by the continuity of $b$
and $c$ that \eqref{th-13} is not possible so that we arrive at a contradiction. Hence we can conclude that $b(t)\ge c(t)$ for all $t\in [T_0,T_n]$.

\vs{6pt}

$(B.4)$ We show that $c$ must be equal to $b$.
For this, let us assume that there exists $t\in[T_0,T_n)$ such that $c(t)<b(t)$. Choose an arbitrary point $x\in (c(t),b(t))$ and
consider the optimal stopping time $\tau^*$ from \eqref{ost-5}  under $\PP_{t,x}$. Inserting $\tau^*$ instead of $s$ in \eqref{th-a} and \eqref{th-5}, and using the optional sampling theorem, we get:
 \begin{align} \label{th-14} \hs{3pc}
&\EE_{t,x} \left[ G(\tau^*,X_{\tau^*})\right]=V(t,x)\\
\label{th-15}&\EE_{t,x} \left[ G(\tau^*,X_{\tau^*})\right]=U^{c}(t,x)+\EE_{t,x} \left[\int_t^{{\tau^*}} H(u,X_u) I\big(X_u \ge c(u)\big)du\right]
\end{align}
where we use that $V(\tau^*,X_{\tau^*})=G(\tau^*,X_{\tau^*})=U^{c}(\tau^*,X_{\tau^*})$ upon recalling that $c\le b$ and $U^{c}=G$ either above $c$ or at $T_n$. Since $U^{c}\le V$ we have from
\eqref{th-14} and \eqref{th-15} that:
 \begin{align} \label{th-16} \hs{3pc}
\EE_{t,x} \left[\int_t^{{\tau^*}} H(u,X_u) I\big(X_u \ge c(u)\big)du\right]\ge 0.
\end{align}
Due to the fact that $H$ is strictly negative above $b$ we see from \eqref{th-16} by continuity of
$b$ and $c$ that such a point $(t,x)$ cannot exist. Thus $c$ must be equal to $b$ and the proof of the theorem is complete.

\end{proof}

\begin{remark}
It can be seen from \eqref{LL} and \eqref{LLL} that the cases of constant and time-varying $\sigma$ have similar numerical complexity for solving the integral equation and computing the value function $V(T_0,\cdot)$.  The difference arises when one is interested in the American swaption price $V^A(t,x)$ at $t<T_0$, which can be computed using \eqref{ost-2}.
Indeed, when only Fourier transform is available, we have to first invert it (see \eqref{pdf}) to obtain the probability density function $p$ and then perform the integration in  \eqref{ost-2}.
\end{remark}
\vs{16pt}

 \section{Pricing problem for fixed-rate receiver}\label{secrec}

In this section we will discuss the problem \eqref{ost-6} corresponding to the fixed-rate receiver. Since minimization problem \eqref{ost-6} has the same payoff function $G$, we will only highlight differences between problems and state the main result (Theorem 4.1). As in the previous section we assume the condition \eqref{assK}.

\vs{6pt}

1.  The continuation and stopping sets are now following
\begin{align} \label{wtC} \hs{5pc}
&\wt{C}^*= \{\, (t,x)\in[T_0,T_n)\! \times\! (0,\infty):\wt{V}(t,x)<G(t,x)\, \} \\[3pt]
 \label{wtD}&\wt{D}^*= \{\, (t,x)\in[T_0,T_n)\! \times\! (0,\infty):\wt{V}(t,x)=G(t,x)\, \}
 \end{align}
and the optimal stopping time in \eqref{ost-6} is given by
\begin{align} \label{OST-2} \hs{5pc}
\tau_b=\inf\ \{\ t\leq s\leq T_n:(s,X_{s})\in \wt{D}^*\ \}.
 \end{align}

Then we can prove the following inequality as in \eqref{b-0}
\begin{align} \label{wtb-1} \hs{5pc}
 \wt{V}(t,x)-\wt{V}(t,y)\ge G(t,x)-G(t,y)
 \end{align}
for $y>x>0$ using the same arguments apart from that $\tau=\tau^*(t,x)$ is now optimal stopping time  for $\wt{V}(t,x)$.
 Now if we take $(t,y)\in \wt{D}^*$, i.e. $\wt{V}(t,y)=G(t,y)$,  we have that $\wt{V}(t,x)=G(t,x)$ and thus $(t,x)\in \wt{D}^*$.
 Therefore we showed that there is a function $\wt{b}:[T_0,T_n)\rightarrow (0,\infty)$ such that
 \begin{align} \label{wtb-2} \hs{5pc}
\wt{D}^*= \{\, (t,x)\in[T_0,T_n)\! \times\! (0,\infty):x\le \wt{b}(t)\, \}.
 \end{align}
Since the problem \eqref{ost-6} is the minimization one, we should not stop when $G>0$ or $H<0$, i.e. we  have that $\wt{b}<g$ and $\wt{b}<h$ on $[T_0,T_n)$, where $g$ and $h$ from \eqref{b-1}. The terminal value of $\wt{b}$ is $\wt{b}(T_n -)=g(T_n -)\vee 0=h(T_n -)\vee 0=\big((\theta \kappa- \alpha(T_n)+ K)/(\alpha(T_n)+\alpha\kappa- K)\big)^+=b(T_n -)$.

\vs{6pt}

2. Standard Markovian arguments lead to the following free-boundary problem (for the value
function $\wt{V}=\wt{V}(t,x)$ and the optimal stopping boundary $\wt{b}=\wt{b}(t)$ to be determined):
\begin{align} \label{PDEb} \hs{5pc}
&\wt{V}_t \p\L_X \wt{V} =0 &\hs{-30pt}\text{in}\;  \wt{C}^*\\
\label{ISb}&\wt{V}(t,\wt{b}(t))=G(t,\wt{b}(t)) &\hs{-30pt}\text{for}\; t\in[T_0,T_n)\\
\label{SFb}&\wt{V}_x (t,\wt{b}(t))=G_x(t,\wt{b}(t)) &\hs{-30pt}\text{for}\; t\in[T_0,T_n) \\
\label{FBP1b}&\wt{V}(t,x)<G(t,x) &\hs{-30pt}\text{in}\; \wt{C}^*\\
\label{FBP2b}&\wt{V}(t,x)=G(t,x) &\hs{-30pt}\text{in}\; \wt{D}^*
\end{align}
where the continuation set $\wt{C}^*$ and the stopping set $\wt{D}^*$ are given by
\begin{align} \label{C-1b} \hs{5pc}
&\wt{C}^*= \{\, (t,x)\in[T_0,T_n)\! \times\! (0,\infty):x>\wt{b}(t)\, \} \\[3pt]
 \label{D-1b}&\wt{D}^*= \{\, (t,x)\in[T_0,T_n)\! \times\! (0,\infty):x\le \wt{b}(t)\, \}.
 \end{align}
It can be shown that this free-boundary problem has a unique solution $\wt{V}$ and $\wt{b}$ which coincide with the
value function \eqref{ost-6} and the optimal stopping boundary respectively.

As for the problem \eqref{ost-5} in previous section,
numerical drawings show that $\wt{b}$ is not monotone function of time (see Figure 3).
It can be intuitively explained as follows: $H_t >0$ on intervals $(T_{m-1},T_m)$ (which is sufficient condition for exhibiting increasing boundary)
however at payment dates $T_m$ the function $H$ exhibits jumps down (which makes the boundary decreasing on the left-side neighborhoods of $T_m$).
\vs{6pt}

3. We now provide the early exercise premium representation formula for the value function $\wt{V}$ which decomposes it into the sum of the expected payoff if we do not exercise until $T_n$ (which is zero) and early exercise premium depending on $\wt{b}$. The optimal stopping boundary $\wt{b}$ will be obtained as the unique solution to the nonlinear integral equation of Volterra type.
We will make use of the following function in Theorem \ref{th:4.1} below
\begin{align} \label{wL} \hs{5pc}
\wt{L}(t,u,x,z)=-\EE_{t,x}\left[ H(u,X_u) I(X_u \le z)\right]
 \end{align}
 for $u\ge t\ge 0$, $x,z>0$, and if $X_u$ has the known probability density function $p$, then
\begin{align} \label{wLL} \hs{5pc}
\wt{L}(t,u,x,z)=-\int_0^z H(u,\wh{x})p(\wh{x};u,x,t)d\wh{x}.
 \end{align}
 Otherwise, we exploit Fourier transform as in the previous section.

 The main result of this section is stated below and provided without proof since it is very similar to Theorem \ref{th:3.1}.
\begin{theorem}\label{th:4.1}
The value function $\wt{V}$ of \eqref{ost-6} has the following representation
\begin{align}\label{th-3} \hs{5pc}
\wt{V}(t,x)=\int_t^{T_n}\wt{L}(t,u,x,\wt{b}(u))du
\end{align}
for $t\in[T_0,T_n)$ and $x\in (0,\infty)$. The optimal stopping boundary $\wt{b}$ in \eqref{ost-6} (see Figure 3) can be characterized as the unique solution to the nonlinear integral equation
\begin{align}\label{th-4} \hs{5pc}
G(t,\wt{b}(t))=\int_t^{T_n} \wt{L}(t,u,\wt{b}(t),\wt{b}(u))du
\end{align}
for $t\in[T_0,T_n)$ in the class of continuous functions
with $\wt{b}(T_n)=\frac{\theta \kappa- \alpha(T_n)+ K}{\alpha(T_n)+\kappa- K}\vee 0$.
\end{theorem}

\begin{figure}[t]
\begin{center}
\includegraphics[scale=0.75]{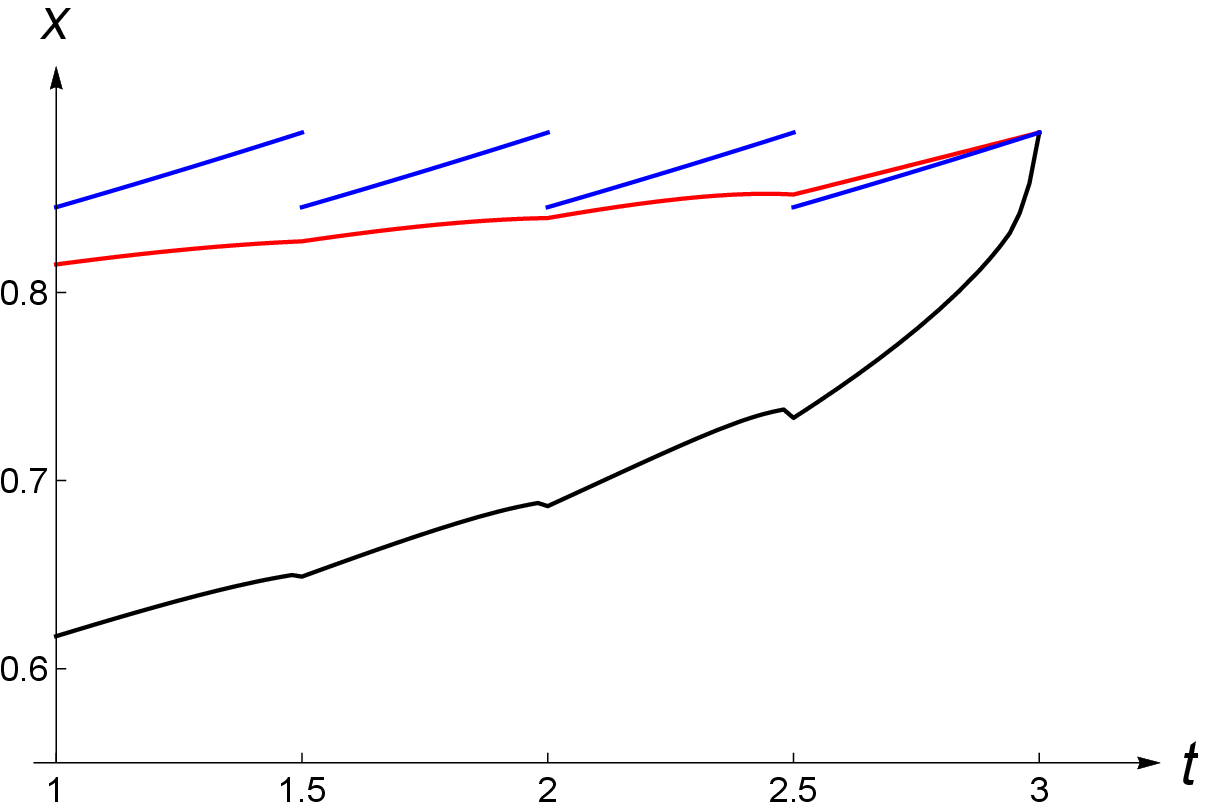}
\end{center}

{\par \leftskip=1.6cm \rightskip=1.6cm \small \ni \vs{-10pt}

\textbf{Figure 3.} A computer drawing of the optimal stopping boundary
$t\mapsto \wt{b}(t)$ for the problem \eqref{ost-6}.
The parameter set is $T_0=1$ year, $\Delta=0.5$ year, $n=4$, $K=0.05$, $\theta=2.55$, $\kappa=0.03$,  $\alpha\equiv \theta\kappa=0.0765$. Coefficient $\sigma$ is time-dependent function and calibrated from European swaption prices.

\par} \vs{10pt}

\end{figure}

\section{Numerical solution to integral equations}\label{secnum}

In this section we provide an algorithm for numerical solution of the integral equations \eqref{th-2} and \eqref{th-4}, and computing the swaption prices \eqref{th-1} and \eqref{th-3}.

In order to obtain the prices of American swaptions \eqref{th-1} and \eqref{th-3} we need to solve numerically integral equations of Volterra type \eqref{th-2} and \eqref{th-4}.
We proved above that $b$ and $\wt{b}$ are unique solutions to the equations \eqref{th-2} and \eqref{th-4}, respectively.
These equations
cannot be solved analytically but can be tackled numerically in an efficient way. The following simple method
can be used to illustrate the latter (see e.g. Chapter 8 in \cite{Detemple}).

Set $t_k=kh$ for $k=0,1,...,N$ where $h=(T_n\m T_0)/N$ so that
the following discrete approximations of the integral equations  \eqref{th-2} and \eqref{th-4}, respectively, are valid:
\begin{align}\label{Rem-1} \hs{5pc}
&G(t_k,b(t_k))=h\sum_{l=k}^{N-1} L\big(t_k,t_{l+1},b(t_k),b(t_{l+1})\big)\\
\label{Rem-2}
&G(t_k,\wt{b}(t_k))=h\sum_{l=k}^{N-1} \wt{L}\left(t_k,t_{l+1},\wt{b}(t_k),\wt{b}(t_{l+1})\right)
\end{align}
for $k= 0,1,...,N\m1$. Setting $k=N\m1$ and $b(t_N)=\wt{b}(t_N)=\frac{\theta \kappa- \alpha(T_n)+ K}{\alpha(T_n)+\kappa- K}\vee 0$ we can solve the equations
\eqref{Rem-1} and \eqref{Rem-2} numerically and get numbers $b(t_{N\m 1})$ and $\wt{b}(t_{N\m 1})$, respectively. Setting $k=N\m2$ and using the values $b(t_{N\m 1})$,
$b(t_{N})$ and $\wt{b}(t_{N\m 1})$, $\wt{b}(t_{N})$, we can solve \eqref{Rem-1} and \eqref{Rem-2} numerically and get numbers $b(t_{N\m 2})$ and $\wt{b}(t_{N\m 2})$, respectively.
Continuing the recursion we obtain $b(t_{N}),b(t_{N-1}),...,b(t_1),b(t_0)$ and
$\wt{b}(t_{N}),\wt{b}(t_{N-1}),...,\wt{b}(t_1),\wt{b}(t_0)$ as approximations of the optimal boundaries $b$ and $\wt{b}$, respectively, at the
points $T_n,T_{n}- h,...,T_0 + h,T_0$ (see Figures 2 and 3 above). We note that we solve  separately the equations for the boundaries $b$ and $\wt{b}$.

Finally, the prices of American swaptions \eqref{th-1} and \eqref{th-3} can be approximated as follows:
\begin{align}\label{Rem-3} \hs{5pc}
&V(t_k,x)=h\sum_{l=k}^{N-1} L\big(t_k,t_{l+1},x,b(t_{l+1})\big)\\
\label{Rem-4}
&\wt{V}(t_k,x)=h\sum_{l=k}^{N-1} \wt{L}\big(t_k,t_{l+1},x,\wt{b}(t_{l+1})\big)
\end{align}
for $k= 0,1,...,N\m 1$ and $x>0$.

\vs{16pt}

\section{Optimal exercise boundaries for swap rates}\label{secoeb}

The formulas \eqref{th-1} and \eqref{th-3} provide the prices of American swaptions for floating-rate receiver and fixed-rate receiver, respectively.
However, the optimal stopping boundaries $b$ and $\wt{b}$ in \eqref{th-2} and \eqref{th-4}
provide the optimal exercise strategies in terms of the latent factor process $X$ that is not directly observable in the financial market in general. Therefore our goal now is to connect the process $X$ with some observable financial object and the natural choice is the \emph{swap rate} of the underlying swap contract.

Let $t\in[T_0,T_n]$ such that $T_{m-1}\le t<T_m$ for some $m=1,...,n$, then let us consider the swap with future payments at $T_m, T_{m+1},....,T_n$.
The swap rate $S_t$ is the fixed rate which makes  $\Pi^{swap}_t=0$ and hence
\begin{align}\label{Rem-5} \hs{5pc}
S_t=\frac{1-P(t,T_n)}{(T_m \m t)P(t,T_m)+\Delta\sum_{j=m+1}^{n} P(t,T_j)}
\end{align}
and by recalling and inserting \eqref{price-2} we get the following relationship between $S_t$ and $X_t$:
\begin{align}\label{Rem-6} \hs{5pc}
S_t=\frac{f_1 (t,X_t)}{f_2 (t,X_t)}=:f(t,X_t)
\end{align}
where
\begin{align}\label{Rem-6a} \hs{-1pc}
f_1 (t,X_t)=&X_t\big[1-e^{-\int_t^{T_n} (\kappa+\alpha(s))ds}\big]+1-e^{-\int_t^{T_n} \alpha(s)ds}(1+\theta)+\theta e^{-\int_t^{T_n} (\kappa+\alpha(s))ds}\\
\label{Rem-6b}
f_2 (t,X_t)=&X_t\left[(T_m \m t)e^{-\int_t^{T_m} (\kappa+\alpha(s))ds}\p\Delta\sum_{j=m+1}^n e^{-\int_t^{T_j} (\kappa+\alpha(s))ds}\right]\\
&\p(T_m \m t)e^{-\int_t^{T_m} \alpha(s)ds}\left(1\p\theta\m\theta e^{-\kappa(T_m-t)}\right)\nonumber\\
&\p\Delta\sum_{j=m+1}^n e^{-\int_t^{T_j} \alpha(s)ds}\left(1\p\theta\m\theta e^{-\kappa(T_j-t)}\right)\nonumber
\end{align}
for $t\in[T_0,T_n]$.

Now if we look into the map $X_t \rightarrow P(t,T;X_t)$ in \eqref{price-2} we see by direct differentiation that  it is strictly decreasing in $X_t$ and therefore
using \eqref{Rem-5} we have that $X_t \rightarrow f(t,X_t)$ is strictly increasing. Thus there is one-to-one relationship between $S_t$ and $X_t$ and
we have that
\begin{align} \label{Rem-7} \hs{5pc}
&x\ge b(t)\qquad \Leftrightarrow \qquad f(t,x)\ge f(t,b(t))\\
\label{Rem-8}
&x\le \wt{b}(t) \qquad \Leftrightarrow  \qquad f(t,x)\le f(t,\wt{b}(t))
 \end{align}
for $t\in[T_0,T_n)$ so that the optimal exercise strategies for fixed-rate payer and fixed-rate receiver in terms of swap rate $S$, respectively, are given as follows
\begin{align} \label{Rem-9} \hs{5pc}
\tau_*=\inf\ \{\ T_0\leq s\leq T_n:S_s \ge R(s)\ \}\\
\label{Rem-10}
\wt{\tau}_*=\inf\ \{\ T_0\leq s\leq T_n:S_s \le \wt{R}(s)\ \}
 \end{align}
where the optimal exercise boundaries $R$ and $\wt{R}$ are given as
\begin{align} \label{Rem-11} \hs{5pc}
R(t)=f(t,b(t))\qquad \wt{R}(t)=f(t,\wt{b}(t))
 \end{align}
 for $t\in[T_0,T_n]$. It can be seen that $R(T_n-)=\wt{R}(T_n-)=K$.
\vs{2pt}

We also note that alternatively one can work with the exercise boundaries related to the short interest rate $r$ using \eqref{price-3} as there is one-to-one relationship between $r$ and $X$ as well.

\vs{16pt}

\section{Bermudan swaption}\label{Bermudan}

In this section, we discuss how the Bermudan swaptions can be priced under the linear-rational framework \eqref{SDE}. The case of
the fixed-rate receiver is symmetric and can be analyzed in the same way.
In contrast with American swaptions, Bermudan contracts can be exercised only at certain number of dates. Usually, these swaptions can be executed
at the maturity $T_0$ of option or at the payment dates of underlying swap, i.e., at $T_i$, $i=1,...,n$.
To keep the valuation as general as possible, we assume that set of possible exercise dates is represented by finite set $\mathcal{T}^B=\{t_j, j=0,...,m\}$ with
$t_0=T_0$ and $t_m=T_n$.
Therefore the Bermudan fixed-rate payer solves the following discrete time optimal stopping problem
\begin{align} \label{berm-ost-1} \hs{6pc}
V^{B}(t,x)=\frac{1}{\zeta_t}\sup_{\tau\in\mathcal{T}^B}\EE_{t,x} \left[G(\tau,X_\tau)\right]
 \end{align}
for $(t,x)\in \mathcal{T}\times(0,\infty)$, where the supremum is taken over all $X$-stopping times $\tau$ with values in the set $\mathcal{T}^B$, and
the function $G$ is given in \eqref{payoff-6}.
Once \eqref{berm-ost-1} is solved, the price $V^B_{t}$ at time $t\in[0,T_0)$
can be then computed as
\begin{align} \label{berm-price-1} \hs{6pc}
V^B(t,x)=\frac{1}{\zeta_t}\EE_{t,x} \left[\zeta_{T_0}V^B(T_0, X_{T_0})\right]
 \end{align}
 using the known distribution of $X_{T_0}$. Obviously, $V^E(t,\cdot)\le V^B(t,\cdot)\le V^A(t,\cdot)$ for any $t\in[0,T_0]$.

 There are at least two standard ways to tackle the discrete time optimal stopping problems: Monte-Carlo simulation and backward induction.
 The former method is more appropriate when the model is not quite tractable or dimension is high. The latter approach is accurate and efficient in low dimensions
 and when the marginal distributions are given. It can be seen below than the linear-rational model allows us to use the backward induction method.
 According to the standard procedure of backward induction(see Chapter 1 in \cite{PS}), the sequence of value functions can be obtained as follows
 \begin{align} \label{berm-price-2} \hs{6pc}
V^{B}(t_j,x)=\max\left(G(t_j,x),\EE_{t_j,x}\left[V^{B}(t_{j+1},X_{t_{j+1}})\right]\right)
 \end{align}
 for $j=0,\ldots,m-1$ and $x>0$ starting from $V^{B}(t_m,\cdot)=V^{B}(T_n,\cdot)=G(T_n,\cdot)=0$.
 The optimal exercise strategy in \eqref{berm-ost-1} starting from $T_0$ is given as
 \begin{align} \label{berm-ost-2} \hs{3pc}
\tau^B_*=\tau^B_*(T_0,x)=\inf\{t_j\in\mathcal{T}^B: V^{B}(t_j,X_{t_j})=G(t_j,X_{t_j}) \}.
 \end{align}

The recursive formulas \eqref{berm-price-2} do not allow for closed form expressions. However, one can perform the following simple numerical scheme. First, we truncate the state space of $X$
by the interval $[0,\bar{X}]$ and discretize it $0=x_0<x_1<...<x_{N}=\bar{X}$ for some $N>0$ and $h_X=x_n\m x_{n-1}$ for $n=1,...,N$. Then we approximate the formula \eqref{berm-price-2} as
 \begin{align} \label{berm-price-3} \hs{3pc}
V^{B}(t_j,x_n)\approx \max\left(G(t_j,x_n),h_X\sum_{i=1}^N V^{B}(t_{j+1},x_i)p(x_i;t_{j+1},x_n,t_j)\right)
 \end{align}
for $j=0,\ldots,m-1$ and $n=1,...,N$. Standard arguments can be applied to show convergence results. If we think of $\mathbf{V^B}(j)\equiv V^{B}(t_j,\cdot)$ as the $N$-dimensional vector and $\mathbf{P}(j)\equiv p(\cdot;t_{j+1},\cdot,t_j)$ as the $N\times N$-matrix for $j=0,\ldots,,m$ , then we can rewrite \eqref{berm-price-3} in the vectorized form. Thus \eqref{berm-price-3} becomes the sequence of recursive vector equations.

We note that in the case of constant $\sigma$, we have a single matrix $\mathbf{P}$ as the process $X$ is time-homogeneous and corresponding probabilities can be obtained from the known density function of non-central chi-square distribution. Then, the algorithm works extremely fast. Otherwise, if $\sigma$ is time-varying, then the computations becomes more involved. First, we have $m$ distinct matrices of transitional probabilities due to time-inhomogeneity of $X$ and secondly we have to invert the Fourier transform. However, if the number $m$ of possible exercise dates is not so large (as we mentioned above, in practice the Bermudan swaption can be exercised twice a year on the same dates as the underlying swap payments occur), then this method it still feasible. Once the family of matrices is computed, the recursive formulas \eqref{berm-price-3} can be easily implemented.  In the next section, we provide the numerical results for the case of time-varying $\sigma(t)$ which we calibrate from the European swaptions data.

\begin{figure}[t]
\begin{center}
\includegraphics[scale=0.7]{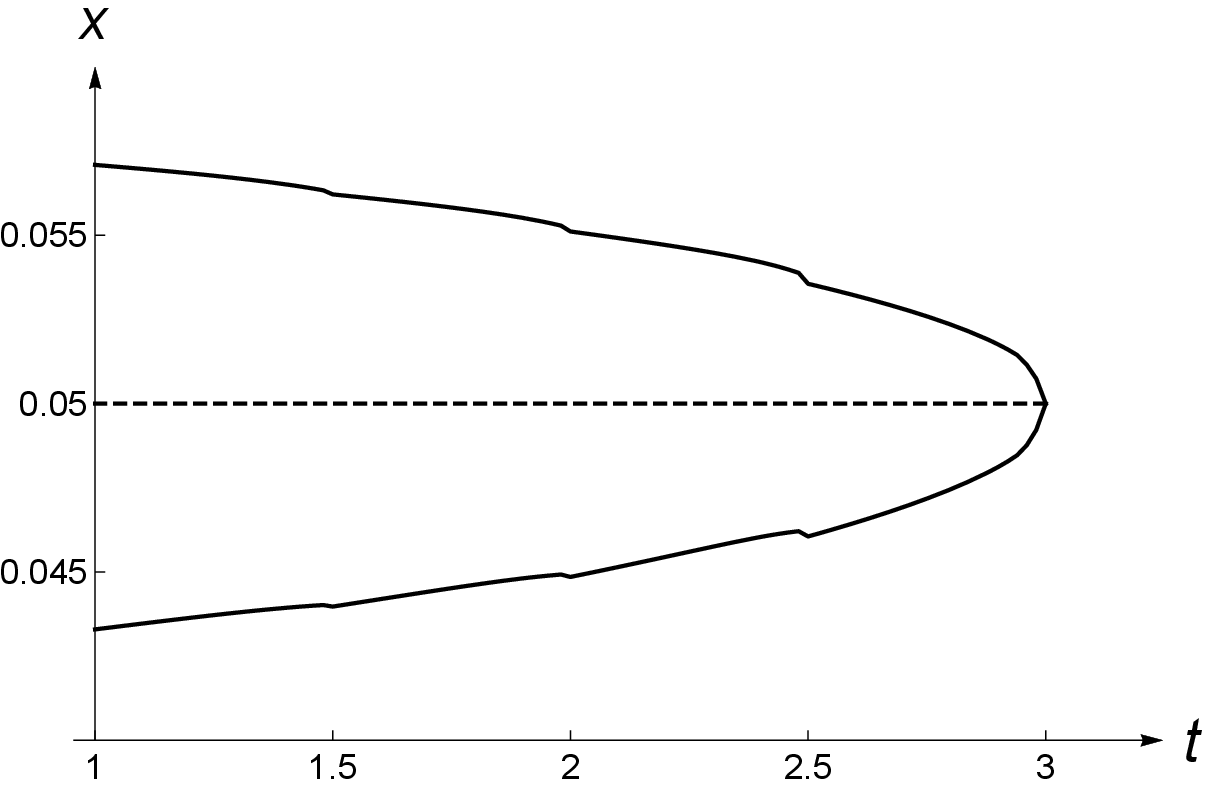}
\end{center}

{\par \leftskip=1.6cm \rightskip=1.6cm \small \ni \vs{-10pt}

\textbf{Figure 4.} A computer drawing of the optimal exercise boundaries $r$ (upper line) and $\wt{r}$ (lower line) in terms of the swap rate $S_t$.
The parameter set is $T_0=1$ year, $\Delta=0.5$ year, $n=4$, $K=0.05$, $\theta=2.55$, $\kappa=0.03$,  $\alpha\equiv \theta\kappa=0.0765$. Coefficient $\sigma$ is time-dependent function and calibrated from European swaption prices.

\par} \vs{10pt}

\end{figure}

\begin{figure}[t]
\begin{center}
\includegraphics[scale=0.7]{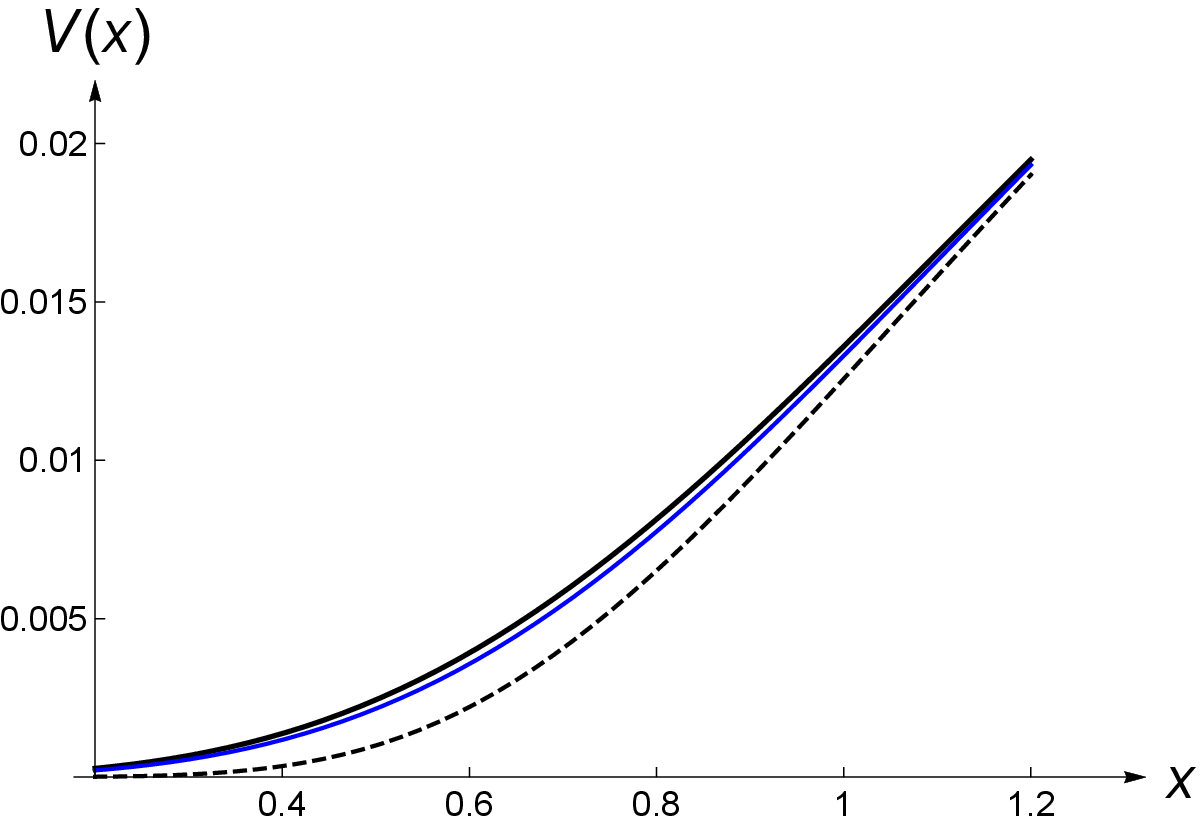}
\end{center}

{\par \leftskip=1.6cm \rightskip=1.6cm \small \ni \vs{-10pt}

\textbf{Figure 5.} The swaption prices of American type $V^A$ (black solid), Bermudan type $V^B$ (blue line) and
European type $V^E$ (dashed) at $t=0$. Bermudan swaption can be exercised at swap payment dates $T_m$ only. The parameter set is $T_0=1$ year, $\Delta=0.5$ year, $n=4$, $K=0.05$, $\theta=2.55$, $\kappa=0.03$,  $\alpha\equiv \theta\kappa=0.0765$. Coefficient $\sigma$ is time-dependent function and calibrated from European swaption prices.

\par} \vs{10pt}

\end{figure}

\vs{16pt}

\section{Numerical results}\label{secnumer}

As an application of Theorems 3.1 and 4.1 with the numerical algorithm described above, we
use the parameters $(\alpha, \kappa, \theta, \sigma(t))$ calibrated to the Euribor swap and swaptions market. Specifically, we set  $\theta=2.55$, $\kappa=0.03$,  $\alpha\equiv \theta\kappa=0.0765$
so that interest rates are bounded below by zero and assumption~\eqref{assK} holds. The coefficient $\sigma$ is a time-dependent function and calibrated from European swaption prices.
The underlying swap starts at $T_0=1$ and has four subsequent semiannual payments, i.e.  $\Delta=0.5$, $n=4$, $T_n=3$.
The nominal value of the swap is 1 million and the fixed rate is $K=0.05$, We are interested in the prices of American, European and Bermudan swaptions at time $t=0$.

We obtain the optimal stopping boundaries $b$ and $\wt{b}$ on $[T_0,T_n]$ (see Figures 2 and 3) as solutions to \eqref{th-2} and \eqref{th-4} using the algorithm above.
We consider the at-the-money swap, i.e.  $S_0=K=0.05$ at $t=0$, then we solve $f(0,X_0)=S_0$ so that the initial factor value is $X_0=0.762$ and
then using \eqref{ost-4} with \eqref{Rem-3} and \eqref{ost-7} with \eqref{Rem-4} we get the American swaption prices $V^A (0,X_0)=0.0072$ and $\wt{V}^A (0,X_0)=0.0068$, respectively, at time $t=0$.
Figure 4 shows a computer drawing of the optimal exercise boundaries $R$ and $\wt{R}$ for swap rate  based on curves $b$ and $\wt{b}$ from Figures 2 and 3.
We observe that the lower boundaries $\wt{b}$ and $\wt{R}$ are not monotone in $t$.

It is remarkable that the boundaries $b$ and $\wt{b}$ are not smooth at the payment dates $T_m$, $m=1,...,n-1$, which is a very rare situation in the optimal stopping theory.
This fact is caused by the discontinuity of $t\mapsto H(t,x)$ at $T_m$, $m=1,...,n-1$, for fixed $x>0$.

Finally, we price the fixed-rate payer European swaption at $t=0$ using \eqref{payoff-3}.
We then consider the fixed-rate payer Bermudan swaption that can be exercised only at underlying swap payment dates. 
Figure 5 shows the values of the fixed-rate payer European, Bermudan and American swaptions at $t=0$ as the functions of $x$. As expected, American price is upper bound for Bermudan swaption price. The prices are close to each other. 
\vs{16pt}

\section{Conclusion}\label{secconc}

The modeling of the state-price density in the linear-rational framework allows us to formulate the American swaption problem as the undiscounted optimal stopping problem for a one-dimensional square-root diffusion process. We characterize the optimal stopping boundaries $b$ and $\wt{b}$ as the unique solution to nonlinear integral equations and using this we obtain the arbitrage-free prices of the American swaptions (see \eqref{th-1} and \eqref{th-3}) and the optimal exercise strategies in terms of swap rates (see \eqref{Rem-9}-\eqref{Rem-10} and Figure 4). The optimal stopping boundaries are not differentiable at payment dates according to numerical solutions, which is a rare situation in the literature on the optimal stopping theory.

Multi-factor factor models tend to empirically outperform one-factor models (see \cite{FLT} for details).
If one considers LRSQ(m,n) specification with $(m+n)$-dimensional factor process $X$, then
the corresponding pricing problem is reduced to the multi-dimensional stopping problem
 \begin{align} \label{ost-3-z} \hs{6pc}
V(t,x)=\sup_{t\le\tau\le T_n}\EE_{t,x} \left[G(\tau,X_\tau)\right]
 \end{align}
for $(t,x)\in [T_0,T_n]\times(0,\infty)^{m+n}$. The function $G$ is affine in factor and is given by
 \begin{align} \label{payoff-1-z} \hs{6pc}
G(t,x)=G^1(t)x_1+...G^{m+n}(t)x_{m+n}+G^0(t)
 \end{align}
 for $(t,x)\in [T_0,T_n]\times(0,\infty)^{m+n}$ for some known functions $G^i$, $i=0,1,...,m+n$.

  We can tackle \eqref{ost-3-z} numerically using, e.g., integral equation approach (for low dimensions), and
backward induction (see Section~\ref{Bermudan}) or Monte-Carlo methods (for higher dimensions). Therefore this leads to an extensive program of
research of American swaptions which we aim to present in subsequent publications.

\vs{12pt}

\begin{center}

\end{center}

\end{document}